\documentclass[11pt,twoside]{article}
\usepackage{asp2010}

\resetcounters

\begin{document}

\title{Revealing the Chamaeleon: First detection of a low-mass stellar halo around the young open cluster $\eta$ Chamaeleontis}
\author{Simon J. Murphy$^{1,3}$, Warrick A. Lawson$^2$ and Michael S. Bessell$^1$}
\affil{$^1$Research School of Astronomy and Astrophyiscs, The Australian National University, Cotter Road, Weston Creek,  ACT 2611, Australia}
\affil{$^2$School of PEMS, University of New South Wales, Australian Defence Force Academy, Canberra, ACT 2600, Australia}
\affil{$^3$\texttt{murphysj@mso.anu.edu.au}}

\begin{abstract}
We have identified several lithium-rich low-mass ($0.08<M<0.3~M_{\odot}$). stars within 5.5 deg of the young open cluster $\eta$ Chamaeleontis, nearly four times the radius of previous search efforts. We propose 4 new probable cluster members and 3 possible members requiring further investigation. Candidates were selected on the basis of DENIS and 2MASS photometry, NOMAD astrometry and extensive follow-up spectroscopy. Several of these stars show substantial variation in their H$\alpha$ emission line strengths on timescales of days to months, with at least one event attributable to accretion from a circumstellar disk. These findings are consistent with a dynamical origin for the current configuration of the cluster, without the need to invoke an abnormally top-heavy Initial Mass Function, as proposed by some authors.
\end{abstract}

\section{Introduction}

The open cluster $\eta$ Chamaeleontis is one of the closest ($d\sim94$~pc) and youngest ($t\sim8$~Myr) stellar aggregates in the Solar neighbourhood. A census of its stellar population currently stands at 18 systems, covering spectral types B8--M5.5 \citep{Mamajek99,Mamajek00,Lawson02,Song04b,Luhman04a,Lyo04,Murphy10}. At high and intermediate masses the cluster Initial Mass Function (IMF) follows that of other star-forming regions and young stellar groups, but there is a clear deficit of members at masses $<$0.15~$M_{\odot}$. Comparing the observed mass function of the cluster to other young groups, \citet{Lyo04} predict that an additional 20 stars and brown dwarfs in the mass range $0.025<M<0.15~M_{\odot}$ remain to be discovered. Efforts to observe this hitherto unseen population have so far failed to find any additional members at either larger radii from the cluster core \citep[][to 1.5~deg, 4 times the radius of known membership]{Luhman04} or to low masses in the cluster core \citep[][to $\sim$13~$M_{\rm Jup}$]{Lyo06}. Failure to find these low--mass members raises a fundamental question: has the cluster's evolution been driven by dynamical interactions which dispersed the stars into a diffuse halo at even larger radii, or does $\eta$~Cha possess an abnormally top-heavy IMF deficient in low--mass objects? The latter result would seemingly be at odds with the growing body of evidence that suggests the IMF is universal and independent of initial star-forming conditions \citep*[for an excellent review see][]{Bastian10}.

\citet*{Moraux07} have attempted to model the observed properties of $\eta$~Cha using $N$-body simulations of the cluster's dynamical evolution starting with standard initial conditions. They are able to replicate the current configuration of the cluster assuming a log-normal IMF and 30--70 initial members. New calculations incorporating binaries (Becker \& Moraux, 2010, in prep.) show almost identical results can be obtained starting with $\sim$20 binary systems. This suggests the deficit of low--mass objects seen in the present day cluster may not be due to a peculiar IMF but to dynamical evolution instead. These simulations predict there should exist a diffuse halo of cluster ejectees beyond the radius currently surveyed. To test the dynamical evolution hypothesis we have undertaken a detailed search for this putative halo of low-mass objects surrounding $\eta$~Cha. Our survey methods and results are described in \citet{Murphy10}, with an extensive follow-up spectroscopy campaign for two candidates thought to be harboring accretion disks presented in \cite{Murphy10b}. Definitive information can be found in these two papers -- in the following contribution we give a precis of our work to date.

\begin{figure}[t] 
   \centering
  \includegraphics[width=0.7\textwidth]{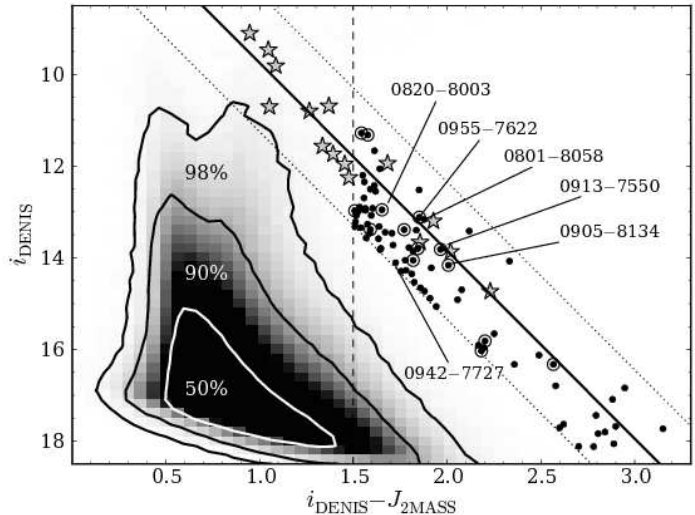}
     \caption{Selection criteria for the 81 photometric candidates from \citet{Murphy10}. Contours show the cumulative total of stars enclosed. Known KM-type cluster members are shown as filled stars. We select candidates (filled circles) within 1.5~mag of the empirical cluster isochrone, having $i-J>1.5$. Proper motion candidates are denoted by open circles. The 6 intermediate-gravity stars are labeled.}
   \label{fig:cmd}
\end{figure}

\section{New members of the low-mass halo surrounding $\eta$ Cha}

\begin{figure}[t] 
   \centering
  \includegraphics[width=0.7\textwidth]{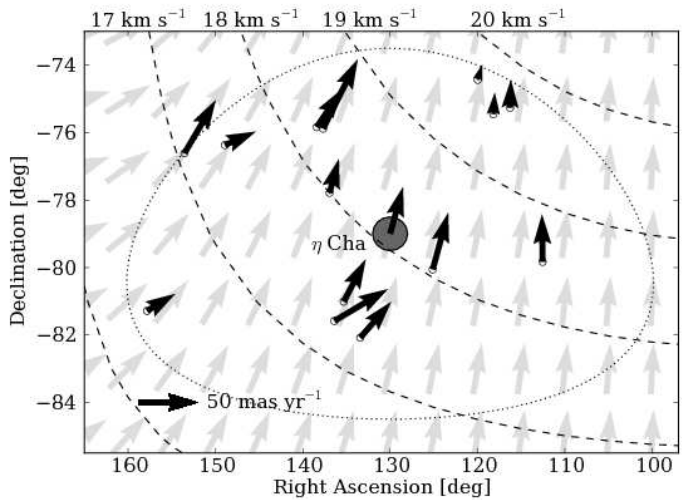}
     \caption{The 14 proper motion candidates and their NOMAD proper motions (black arrows), compared to the cluster space motion projected onto the plane of the sky (gray arrows). Complementary radial velocity contours are shown by dashed lines. The 5.5~deg radius of our survey is given by the dotted border.}
   \label{fig:pm}
\end{figure}

From 2MASS and DENIS photometry of the $1.2\times10^6$ sources within 5.5~deg of $\eta$~Cha we selected 81 photometric candidates having $i_\textrm{\tiny{DENIS}}$ photometry within 1.5~mag of the empirical isochrone of confirmed $\eta$~Cha members from the literature (Figure~\ref{fig:cmd}). Fourteen photometric candidates have NOMAD proper motions consistent with cluster membership. Na\"{i}vely one could expect to find proper motion candidates simply by selecting objects with proper motions similar to the cluster mean. This would however bias any survey to candidates near the cluster on the sky. Figure~\ref{fig:pm} demonstrates the difficultly in using proper motion selection over such a large survey area. For a given $UVW$ space motion the resultant proper motion vector (and radial velocity) depends on both sky position and distance. Given the large angular extent of the survey area, this can have a substantial effect on the expected proper motions and radial velocities of our candidates. We have corrected for this effect by comparing the observed proper motions to those expected at the position of each photometric candidate. The 14 resulting proper motion candidates are shown in Figure~\ref{fig:pm}. Of these stars, four have intermediate gravities, as measured from the strength of the Na~\textsc{i} 8200~\AA\ absorption doublet. An intermediate gravity between dwarfs and giants is characteristic of Pre-Main Sequence (PMS) stars, which are still contracting towards their Main Sequence radii. Two other stars have intermediate gravities but are not proper motion members. All six stars have Li~\textsc{i} 6708~\AA\ equivalent widths consistent with a 5--10~Myr PMS population. In addition to the 2MASS/DENIS sources we also investigated a selection of stars from the literature. In particular several of the ROSAT-selected PMS stars from \citet{Covino97} met the photometric, proper motion and Lithium requirements and were added to our final candidate list for dynamical modeling.

\subsection{Dynamical modeling}

\begin{figure}[p] 
   \centering
  \includegraphics[width=0.6\textwidth]{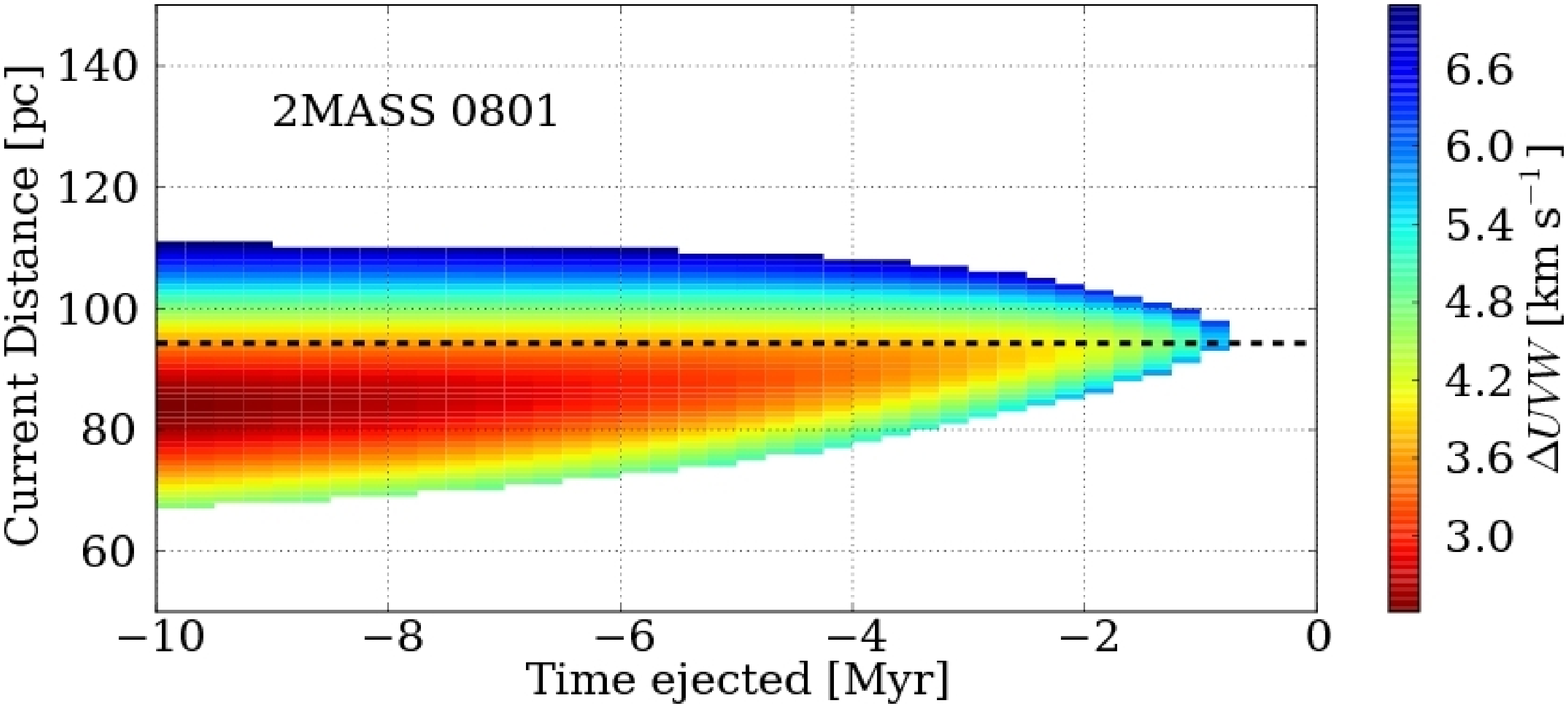}
  \includegraphics[width=0.6\textwidth]{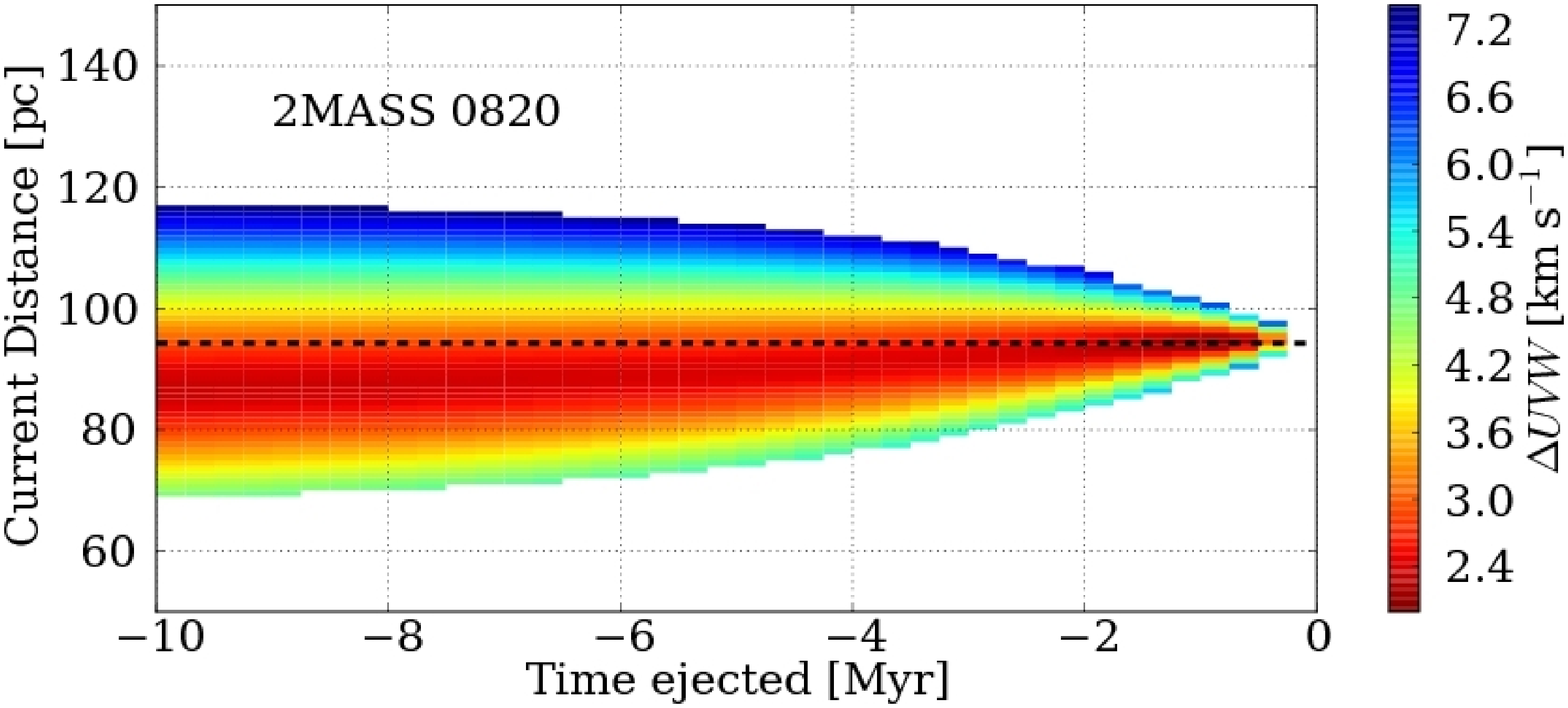}
  \includegraphics[width=0.6\textwidth]{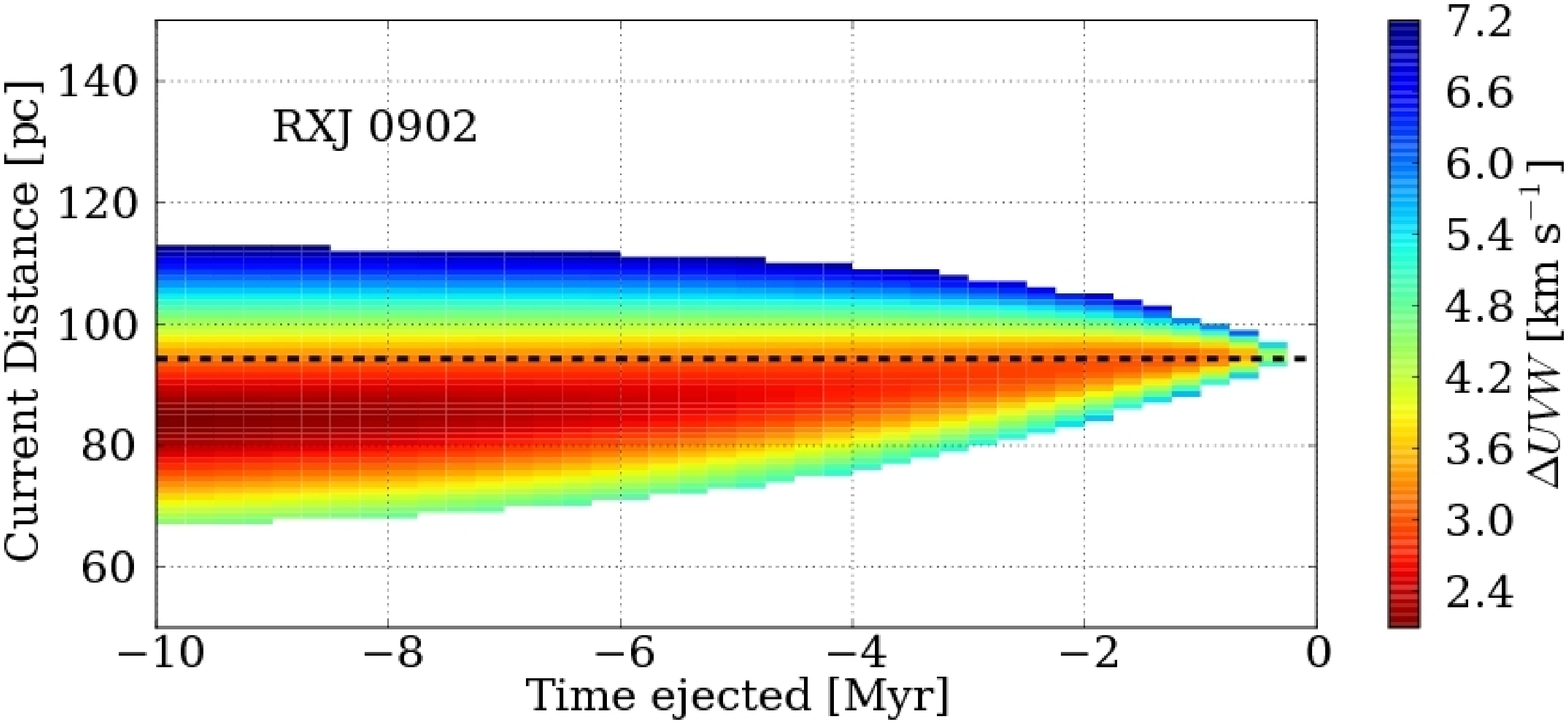}
    \includegraphics[width=0.6\textwidth]{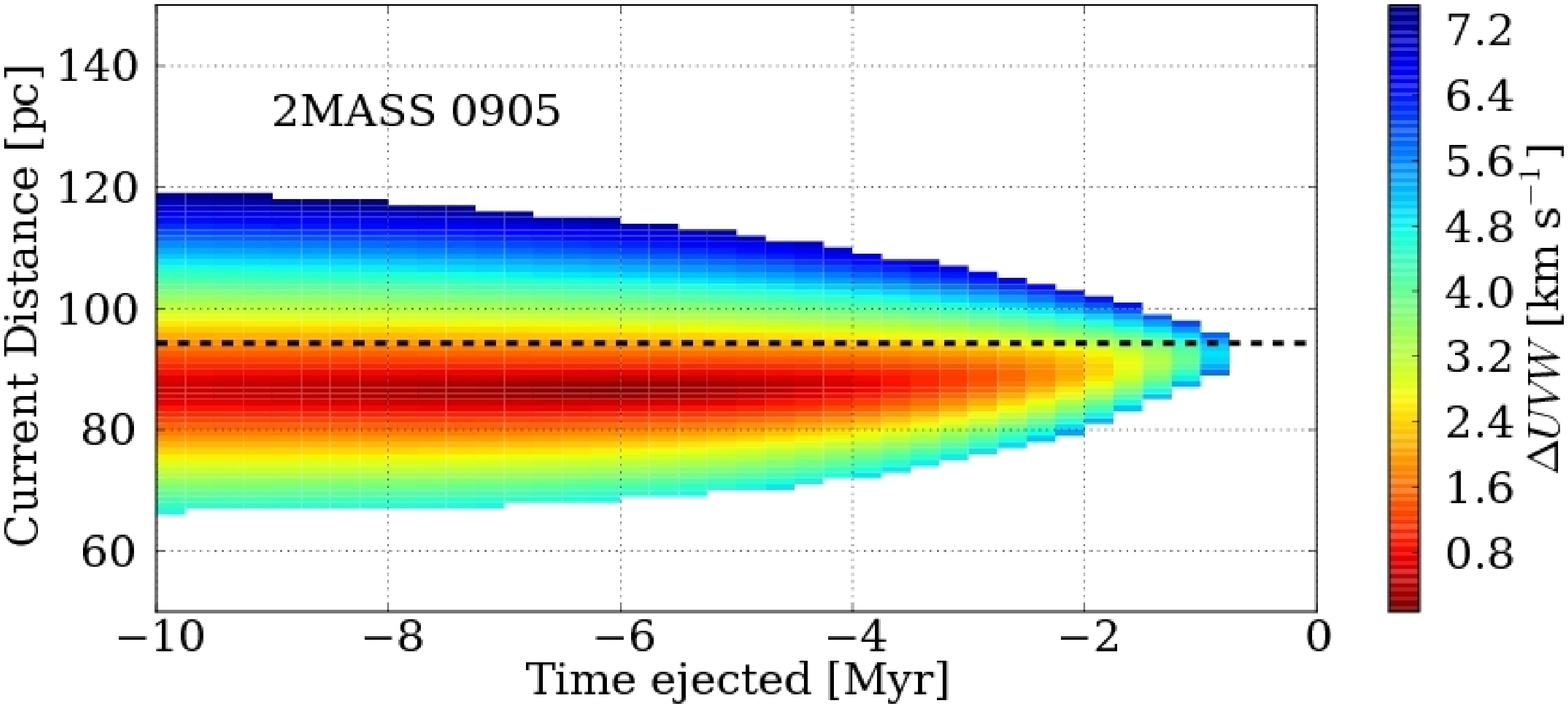}
    \includegraphics[width=0.6\textwidth]{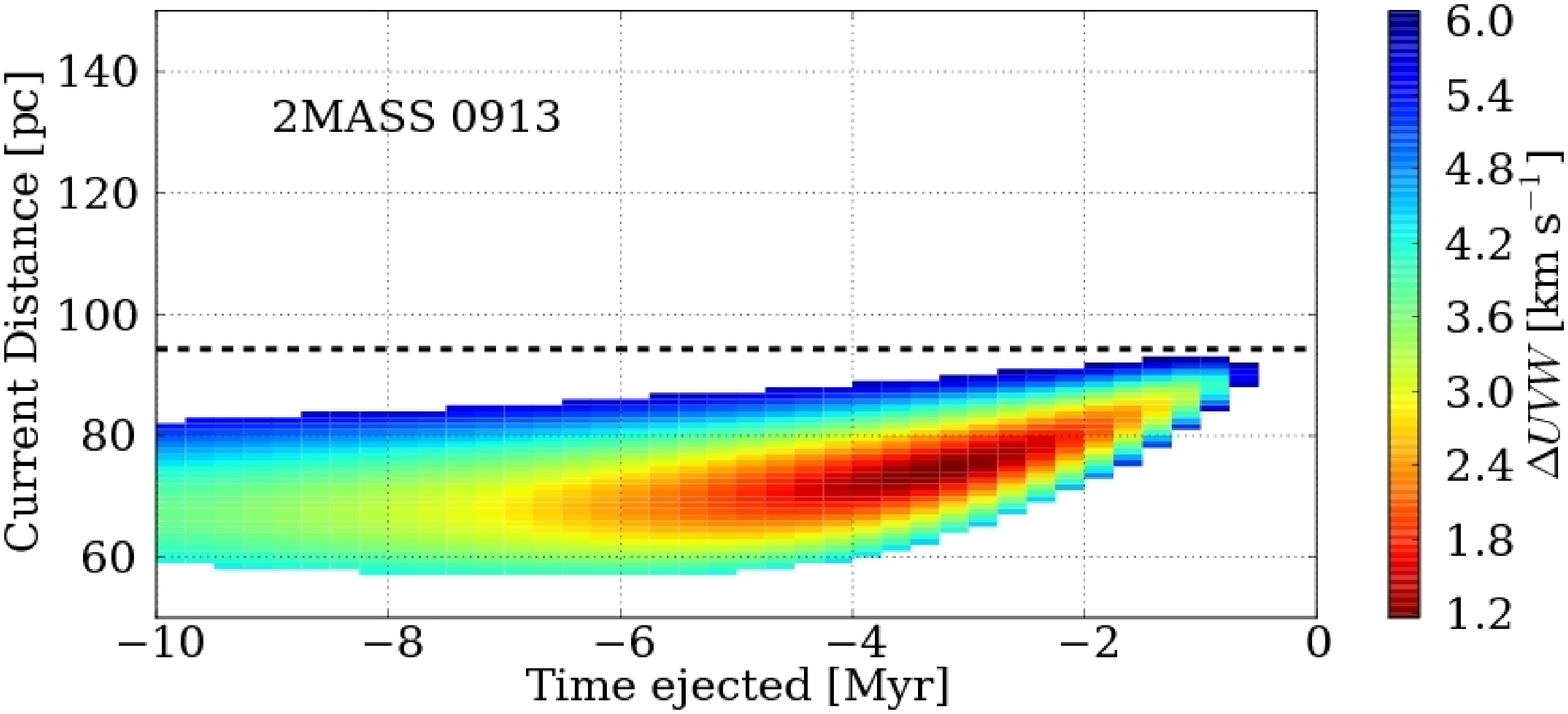}
     \caption{Results of the dynamical modeling for the 4 new probable members and the possible accretor 2MASS~0801. Based on these plots (and similar ones for ejection speed) we can estimate when and how fast the stars were ejected from the cluster and at what distances they currently lie. More information may be found in  \citet{Murphy10}.}
   \label{fig:deltav}
\end{figure}

If the remaining candidates are in fact ejected members of $\eta$~Cha we do not expect their space motions to be identical to that of the cluster proper. They will have each been imparted some ejection velocity which acts over time to disperse the star away from the cluster core. To determine the possible epoch and magnitude of this impulse we modeled the space motion of the candidates as a function of ejection time and current distance. Figure~\ref{fig:deltav} shows the results of these simulations for several candidates. Similar plots for the other candidiates and a more detailed explanation of the method used are presented in \citet{Murphy10}. The colour-map in Figure~\ref{fig:deltav} indicates the difference between the observed space motion and that expected from the current position of the star assuming ejection from the cluster. A similar map can be made for ejection speed. The kinematic distances estimated from the modeling can be checked against those expected from a star's position in the cluster CMD.  Overall there is excellent agreement between the kinematic and photometric distances of the candidates. From such comparisons we have identified 4 new probable and 3 possible outlying members of $\eta$~Cha, at radii of 1.5--5~deg (2.5--8~pc).

\subsection{Dynamical evolution or abnormal IMF?}

Are these numbers consistent with a dynamical origin for the paucity of low-mass objects currently observed in the cluster? To answer this we consider the radial distribution of ejectees presented in \citet{Moraux07}. Integrating the distribution out to 9~pc (5.5 deg) we could expect to find up to 6 stars across all masses in the survey area. Our blue photometry limit of $i-J>1.5$ corresponds to an approximate spectral type of M3, whereas our spectroscopic campaign is complete to  $i-J<2$, approximately M5. Transforming these spectral types to temperatures and then to masses from an 8~Myr \citet{Baraffe98} isochrone gives an \emph{approximate} surveyed mass range of $0.08<M<0.3~M_{\odot}$. \citeauthor{Moraux07}\ find the mass distribution of ejectees is roughly constant with radius and consistent with the \citet{Chabrier03} input IMF. We therefore integrate this IMF over the above mass range to find the fraction of stars at those masses, this is $\sim$40\%. Hence we can expect to find 2--3 bona fide new $\eta$~Cha members within the surveyed area and mass range. Our discovery of 3 probable 2MASS/DENIS members is consistent with this prediction. \emph{We can therefore conclude that dynamical evolution is solely responsible for the current configuration of $\eta$~Cha and it is not necessary to invoke an IMF deficient in low--mass objects. }

\section{Episodic disk accretion at $\sim$8~Myr}

Nearby, isolated groups such as $\eta$ Cha are ideal laboratories for investigating the dynamical evolution of young star clusters, in particular the influence that dynamics have had on the evolution of protoplanetary disks. Young clusters show a steady decline in the number of stars having disks and signatures of accretion with age \citep{Mohanty05,Jayawardhana06}. By an age of $\sim$5~Myr, 90--95\% of all young cluster members have stopped accreting material at a significant rate, yet $\sim$20\% of objects retain enough dust in their disks to produce a mid-IR excess \citep{Fedele10}. By investigating the disk and accretion properties of any new dispersed members of  $\eta$~Cha, we can hope to gain a more unbiased view of the cluster as a whole, as well as addressing any influence dynamical interactions have had on disk evolution. For instance, we might expect any disks surrounding the outlying members to be truncated in mass and radius by the strong dynamical forces responsible for their dispersal, or show other systematic differences when compared to the evolution of the dynamically less-evolved inner members.

\begin{figure}[t] 
   \centering
 \plottwo{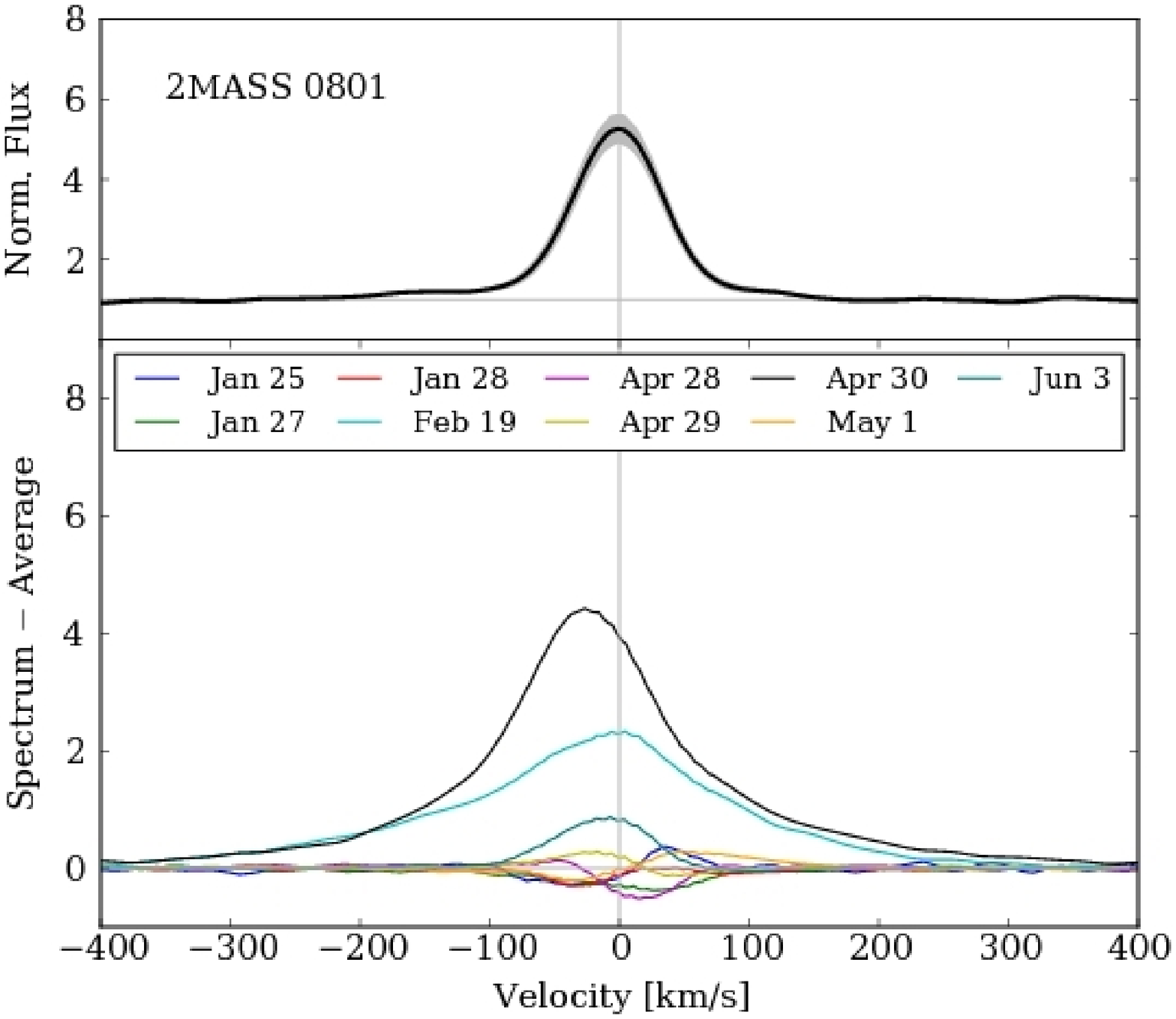}{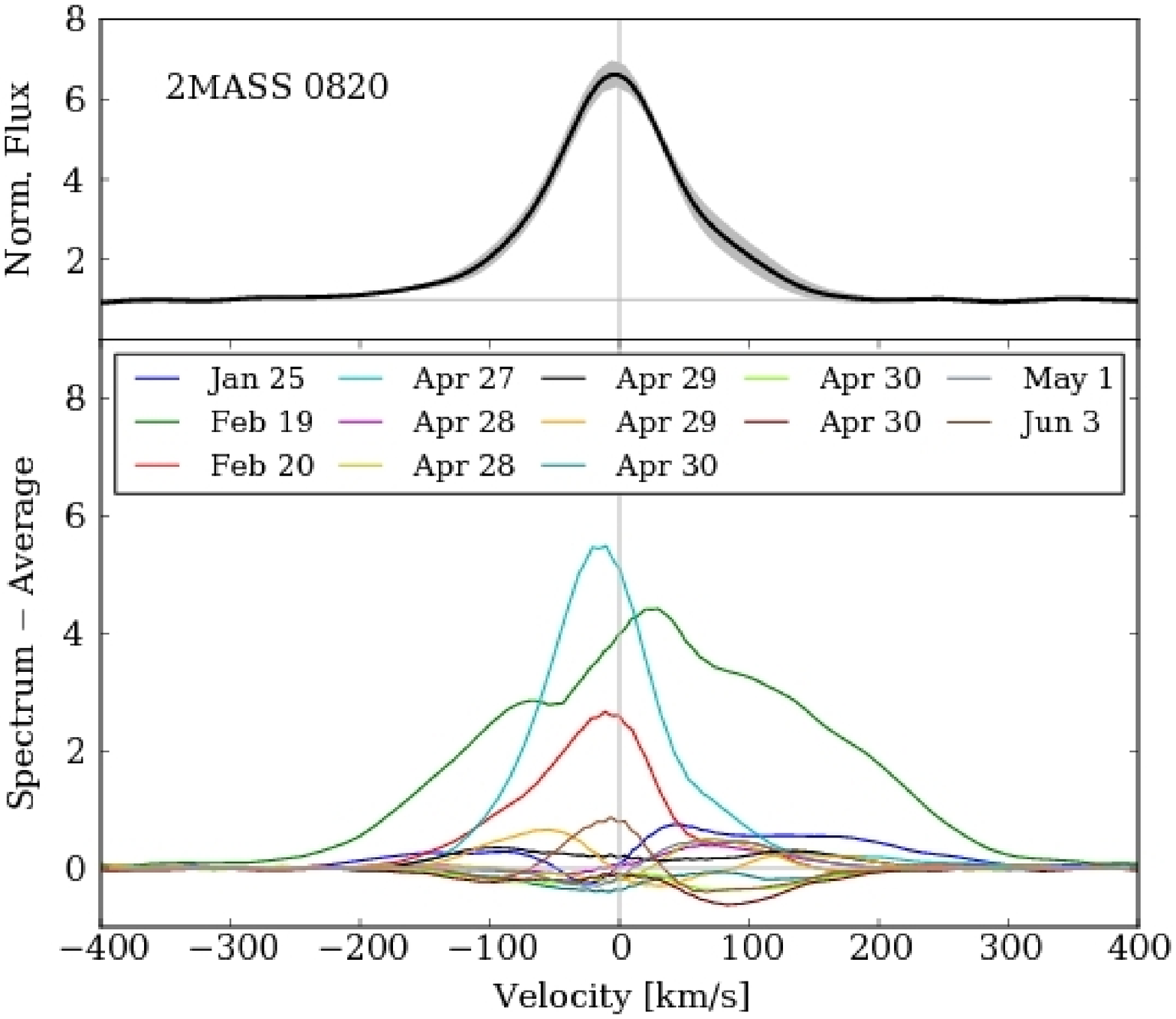}
      \caption{H$\alpha$ velocity profiles for the 9 observations of 2MASS~0801 (left panel) and 13 observations of 2MASS~0820. In each plot the top panel shows the average quiescent spectrum spectrum and the standard deviation of quiescent spectra around the mean (shaded region). The bottom panel shows the variation around the mean for all epochs. Note the broad, multicomponent residual on Feb 19 for 2MASS~0820.}
   \label{fig:halpha}
\end{figure}

Two of our candidates showed strong and highly variable H$\alpha$ emission over the six months of observations: the probable candidate 2MASS~0820--8003 and the possible candidate 2MASS~0801--8058. This star is a possible binary based on its elevated position in the cluster CMD and unusually large radial velocity variations. We obtained 9 medium-resolution ($R\approx7000$) observations of 2MASS~0801 over 2010~January--June and 13 epochs for 2MASS~0820. In addition to strong H$\alpha$ emission both stars recurrently showed He~\textsc{i} 5876~\AA, 6678~\AA\, and Na~\textsc{i}~D emission, often associated with accretion. Our full results can be found in \citet{Murphy10b}, including the variation of the H$\alpha$ line in the Equivalent Width--10\% Intensity Velocity Width plane \citep{Lawson04,Jayawardhana06} and comparisons with other multi-epoch studies of $\eta$~Cha members. We present here only the variation in the H$\alpha$ line profile shapes (Figure \ref{fig:halpha}). Immediately apparent are the broad residual profiles, tracing velocities up to $\pm$200--300~km~s$^{-1}$. The February 19 epoch of 2MASS~0820 shows a residual velocity profile reaching to $\pm300$~km~s$^{-1}$, with four distinct components visible and a large red asymmetry. Velocity shifts in the peaks of individual residual spectra are present at up to several tens of km~s$^{-1}$.

 Two possible mechanisms are usually invoked to explain such distinctive line profiles -- accretion from a circumstellar disk or chromospheric activity. The work of \citet{Montes98} has shown that in some older T Tauri stars the H$\alpha$ line profile cannot be fitted by a single gaussian and two components are necessary: a narrow gaussian of FWHM $<$100~km~s$^{-1}$ and a much broader component with FWHM 100--500~km~s$^{-1}$, sometimes offset in wavelength from the narrow component. They attribute these line profiles to chromospheric micro-flaring. Micro-flares are frequent, short duration events and have large-scale motions that could explain the broad wings observed in the lines and the residual spectra in Figure~\ref{fig:halpha}. Both stars also show He~\textsc{i} 6678~\AA\ in emission when H$\alpha$ levels are strongest. Strong He~\textsc{i} 6678~\AA\ emission is an accretion diagnostic as it is only present in low-levels ($\ll$1~\AA) in older chromospherically active stars. While we do detect strong (1.5~\AA) emission in the April~30 outburst spectrum of 2MASS~0801, at all other epochs where we detect the line it is weak ($\sim$0.5~\AA). Given the weak He~\textsc{i} line strengths generally observed in our stars and the simple gaussian-like profiles of the residual spectra we do not have strong evidence for ongoing accretion. Chromospheric activity is a much more likely explanation for the observed line profiles. Only the February~19 spectrum of 2MASS~0820 shows a broad, asymmetric residual characteristic of accretion. Multiple components are present at velocities up to $\pm$300~km~s$^{-1}$, presumably tracing the ballistic infall of material from the inner edge of the disk onto the stellar surface.

Our results show that H$\alpha$ variability in $\sim$8~Myr PMS stars can be substantial on both short (hours--days) and long (months) timescales. This variation is probably driven primarily by chromospheric activity, which can generate broad H$\alpha$ profiles mimicking accretion over short timescales. However we also have evidence for at least one accretion event in 2MASS~0820 which requires follow-up observations. Additional mid-IR observations will be necessary to detect the presence of any circumstellar disk around the star feeding the accretion. Assuming the duty-cycle of episodic accretion is low, single-epoch surveys of accreting objects, especially in the critical age range 5--10~Myr when inner disks are being cleared and giant planet formation takes place, are likely missing a large fraction of accreting objects. Gas depletion timescales derived from the fraction of accretors are therefore likely underestimated. A larger survey of the disk and accretion properties of outlying $\eta$~Cha members, combined with more detailed investigation of the true accreting fraction of PMS clusters from multi-epoch surveys is needed to resolve the issue.

\acknowledgements SJM acknowledges the generous support of the LOC and the receipt of an RSAA Alex Rodgers Traveling Scholarship and Joan Duffield Research Scholarship.

\end{document}